# Minimalist design of polymer-oligopeptide hybrid as intrinsically disordered protein-mimicking scaffold for artificial membraneless organelle

Jianhui Liu [1], Fariza Zhorabek [1], Xin Dai [2], Jinqing Huang [2] and Ying Chau [1,*]

Affiliations:

1. Department of Chemical and Biological Engineering, the Hong Kong University of Science and Technology, Clear Water Bay, Kowloon, Hong Kong.

2. Department of Chemistry, the Hong Kong University of Science and Technology, Clear Water Bay, Kowloon, Hong Kong.

**Abstract:**

Liquid-liquid phase separation (LLPS) is an emerging and universal mechanism for intracellular biomolecule organization, particularly, *via* the formation of membraneless organelles (MOs). Intrinsically disordered proteins (IDPs) are the main constituents of MOs, wherein multivalent interactions and low-complexity domains (LCDs) drive LLPS. Using short oligopeptide derived from LCDs as 'stickers' and dextran backbones as 'spacers', we designed polymer-oligopeptide hybrids to mimic the multivalent FUS protein as represented by the 'stickers-and-spacers' model. We demonstrated that hybrids underwent LLPS and self-assembled into micron-sized (mostly 1−10 μm, resembling LLPS in vitro and in living cells) compartments displaying liquid-like properties. Furthermore, the droplets formed were capable of recruiting proteins and RNAs, whilst providing a favorable environment for enhanced biochemical reaction, thereby mimicking the function of natural MOs. We envision this simple yet versatile model system will help elucidate the molecular interactions implicated in MO formation and pave ways to a new type of biomimetic materials.

# Introduction

Cellular metabolism requires precise spatiotemporal regulation of numerous biomolecules. Besides lipid bilayer membrane-delimited compartmentalization[1], membraneless organelles (MOs) formed by liquid-liquid phase separation (LLPS) provide another universal intracellular organization. MOs have aroused intense interests from multidisciplinary scientists owing to the ubiquitous biological implications in cellular physiology and disease[2,3]. In contrast to the thermodynamically stable amyloid-like structures[4], MO structures are dynamic and reversible[5]. Most MOs contain IDPs harboring low-complexity domains (LCDs)[6], which are responsible for driving LLPS *via* weak and multivalent interactions. Reported in natural living cells[7,8] and reconstituted systems in vitro[9–13], large LCDs[8,10,14] (usually over 100 residues) or LCD-harboring engineered proteins[9–11] are the building blocks for LLPS. Notwithstanding, the minimal structural motifs of large LCDs, sequence determinants, and the molecular arrangement required for LLPS are yet to be revealed.

The intricate molecular interactions implicated in IDPs can be depicted using a simplified 'stickers-and-spacers'[15] framework derived from Flory-Huggins theory, wherein the mean-field free energy was enhanced from 'sticker' interactions[16]. Modules driving molecular attractions are considered as 'stickers', whilst modules providing flexible linkage (of 'stickers') with no significant attractions are considered as 'spacers[16]'. We reason 'stickers-and-spacers[16]' interaction mode with prominent multivalency[10,17,18] are the key and universal molecular determinants of LLPS of IDPs. Hence, we are inspired to employ a bottom-up and minimalist approach to design biomimetics of the scaffolding proteins of MOs, which we term "IDP-mimicking polymer-oligopeptide hybrid" (IPH). We target a simplified model system with concise and well-defined interaction modules to help elucidate biological LLPS from the molecular structure level, as well as synthesize artificial membraneless organelles (AMOs) and evaluate properties thereof. IPH was chemically synthesized by grafting hydrophilic, flexible polymer chains (as the 'spacers') with weakly interacting, LCD-derived, short oligopeptides (as the 'stickers'). The key molecular characteristics, including molecular weights, patterning and composition of structural motifs were chosen to mimic natural IDPs. We employed turbidimetry and optical microscopy for characterizing micron-sized liquid droplets formation, as well as fluorescence recovery after photobleaching (FRAP) for characterizing the internal mobility. The LLPS behavior was further investigated for IPH with modulated structural parameters, namely, degree of modification[19] (DM) of peptide, molecular weight (MW) and tyrosine/arginine ratio (Y/R).

MOs are hubs for numerous intricate biochemical reactions, owing to the coexistence of compartmentalized structures and high dynamics of molecules[2]. Generally hosting protein and RNA, MOs are ubiquitous in both cytoplasm[20] and nucleus[21], widely implicated in multifaceted RNA metabolism[22]. We thus evaluate the hypothesis that AMOs can harbor certain functions of natural MOs, namely, preferential and reversible recruitment of cargo proteins, RNA molecules and enhancement of the associated biochemical reactions.

## Results and discussion

**FUS-mimicking IPH forms droplets *via* LLPS in vitro**

As a general structural feature, IDPs contain binding elements with high valency and modest affinity, between which long and flexible linkers are interspersed[23]. We hypothesize IDP-mimicking hybrid materials with multivalent weak molecular interactions could undergo LLPS to form MO-like compartments. We designed and synthesized IPH *via* grafting CLCDP/CRP dual peptides to vinylsulfone-modified dextran (Dex-VS), which is aimed to recapitulate the structural features of FUS, a representative IDP (Fig. 1a). We choose charge-neutral dextran as the backbone material because of its hydrophilicity, good biocompatibility, and disordered and flexible random-coil-like nature of the polymeric chains[24], which are essential molecular features of 'spacer' modules of IDPs. The hydroxyl groups on the dextran backbone were functionalized with thiol-reactive vinylsulfone (VS) groups[19] to provide chemical anchorage points for oligopeptides, and $^1$H-NMR was applied to quantify the DMs (Supplementary Fig. 1b, c).

Weak and reversible molecular interactions are prevalent in LCDs and pivotal for driving LLPS[25]. Recently, short (generally < 10 residues) reversible amyloid cores (RACs), the low-complexity motifs of RNA-binding IDPs including FUS[26] (Supplementary Fig. 2a), TDP-43[27] and hnRNP family[28], were revealed to form labile and reversible fibrils reminiscent of structures formed by full-length IDPs, thereby supporting that RACs are drivers of intra- and inter-molecular interactions. Moreover, RACs harbor aromatic residues that stabilize weak molecular interactions and labile self-assemblies[25]. We thus sought to leverage RACs as the minimized structural motif units as the 'sticker' modules for recapitulating the LLPS behavior of full-length IDPs[29–31]. Specifically, inspired by FUS protein, we designed and synthesized two peptides: cysteine-terminated LCD-like peptide (CLCDP) and cysteine-terminated (arginine-glycine-glycine)-containing peptide (CRP) (Supplementary Fig. 2d, e). CLCDP contains a flexible segment (CGG) conjugated to an aromatic-rich RAC segment (SYSGYS). CRP contains a flexible segment (CGG), linked to repeats of RGG. RGG repeats are included for three reasons: the abundance of RGG segment in FUS protein[8], the dominance of cation-π-type interaction from tyrosine-arginine (Y-R) pairs in IDPs[8,18,29,32,33], and the presence of a positive charge for nucleic acids recruitment[34,35].

In the design of IPH, we also considered three molecular features of FUS protein, namely, the size of macromolecules, the Y/R ratio, and the patterning of RACs (Supplementary Fig. 2b).

One construct, termed IPH*, was designed (Supplementary Fig. 2 a-c) and synthesized (Supplementary Fig. 3a) to mimic the three aspects to the most extent. As the molecular module sufficient for driving LLPS[8], the LCD of FUS, harboring 214 residues, is mimicked by selecting 40 kDa dextran backbone with 247 repeating units. As an intrinsic parameter of IDPs, governing LLPS propensity[8], the Y/R ratio of FUS protein, equaling 0.973, is mimicked by using IPH with comparable Y/R ratio (Y/R=1.03). The DM of CLCDP (13.3 %) was chosen to allow the spacing of oligopeptide stickers to mimic the patterning of RACs in the low-complexity domains of FUS protein, which is slightly higher than designer DM (9.35 %, Supplementary Fig. 2b and Supplementary Methods) to ensure sufficient driving force of LLPS.

IPH* underwent LLPS to form micron-sized droplets under physiological conditions in vitro (Fig. 1b), reminiscent of LLPS of parental full-length FUS protein in living cells[5,36] and in vitro[32]. The liquid-like nature of droplets was confirmed by wetting phenomenon, fusion event and FRAP (Fig. 1c-e and Supplementary Video 1). The droplets formed by IPH* allowed rapid material rearrangement with apparent diffusivity $D_{app} = 0.0114\ \mu m^2/s$, which is within the range found for in vitro LLPS system constructed by FUS protein[37] ($D_{app} = 0.002 - 0.016\ \mu m^2/s$) and LAF-1 protein[18] ($D_{app} = 0.025 - 0.01\ \mu m^2/s$). The propensity of LLPS was evaluated by the density and size of droplets formed under microscopy, as well as turbidimetry. The extent of LLPS of IPH* depended on the ionic strength (I) and pH conditions (Fig. 1f-h and Supplementary Fig. 4c). Interestingly, LLPS exhibited highest propensity at physiological ionic strength ([NaCl]=150 mM), whilst LLPS is less sensitive to pH change across the broad range tested. The ionic strength-and-pH phase diagram exhibits 'diagonal-like' phase behavior, namely, LLPS is more prone to happen in the two quadrants: low-I-high-pH (LIHP) and high-I-low-pH (HILP) (Fig. 1f and Supplementary Fig. 4c), in which arginines are moderately charged. In the other two quadrants, low-I-low-pH (LILP) and 'high-I-high-pH (HIHP), which correspond to strong and weak charged states of arginines, respectively, LLPS is less favored. The phase behavior is determined by the collective contribution of promiscuous interactions including cation-π, π-π and cation-cation interactions. We reason that cation-π interactions are more important than π-π interactions for driving LLPS, while cation-cation interactions hamper LLPS in our system. This explains why arginine sites are moderately charged in the favorable quadrants. For LILP, arginine sites are strongly charged and generate strong counteractive repulsive interactions. For HIHP, arginine sites are largely deprotonated and prevent cation-π interactions. These observations accord with the cation-π-type tyrosine-arginine interactions being indispensable for IDPs to undergo LLPS[8]. IPH* phase-separated in the absence and presence of PEG (as a crowding agent) in a concentration-dependent fashion (Fig. 1i and Supplementary Fig. 4a). Low-dose incorporation of crowding agent could enhance the propensity of phase separation in a dose-dependent manner, whilst high-dose incorporation could undermine the propensity thereof (Fig. 1j and Supplementary Fig. 4b). The upper critical solution temperature (UCST) behavior, i.e., higher propensity to phase separate under lower temperature, was characterized by temperature-dependent turbidimetry assay (Fig. 1k) and optical microscopy (Supplementary Fig. 4e) under physiologically relevant concentrations, whereas LLPS gradually evolves into dispersed solution upon heating across the 35 °C to 45 °C regime, reminiscent of the UCST behavior of FUS[38] and some other IDPs[10,39]. We employ 1,6-hexanediol (HDO) assay to test the metastability and reversibility of IPH droplets. Generally, assemblies of liquid-like and labile nature can be disrupted by HDO, while strong assemblies such as amyloid plaques cannot[40–43]. The dose-dependent disruption of droplets and droplet recovery after HDO removal were confirmed by both turbidimetry assay (Fig. 1l) and optical microscopy (Supplementary Fig. 4d), supporting the metastability and reversibility of IPH droplets, respectively.

**LLPS behavior is dependent on the molecular property and structure of IPH**

Valency[10,44] and 'sticker-sticker' affinity[8,44] of IDPs have been shown as the molecular determinants of multivalent interactions driving LLPS. We thus sought to investigate specific molecular determinants of LLPS for the IPH system. IPHs varying in DM, MW and Y/R ratio were synthesized (Supplementary Fig. 3b-d). For a systematic study, only one parameter was changed while the other two remained the same as those for IPH*. Based on the 'stickers-and-spacers' model, we hypothesize that DM and MW will affect the valency of interactions, while the Y/R ratio will influence the strength of 'sticker-sticker' interactions.

Firstly, we investigated the effect of DM by oligopeptides on phase behavior. Only IPH with DM higher than a threshold DM (22.3%<$DM_{threshold}$<37.9%) could exibit LLPS under physiological conditions in vitro, and LLPS propensity increased with increasing DM (37.9%, 50.2% and 91.7%) (Fig. 2a and Supplementary Fig. 5a, b). This is consistent with the strong dependence of the phase behavior on the valency of the 'stickers' of IDPs, that is, higher valency allows the formation of LLPS at lower IDP concentration[10]. All high-DM IPHs were prone to phase separate under moderate ionic strength, and in particular, maximum turbidity was observed under the physiological value (150 mM NaCl) (Fig. 2b and Supplementary Fig. 5c). IPHs with lower DMs were prone to phase separate under alkaline conditions, and LLPS propensity at acidic pH increased drastically with increasing DM (Fig. 2c and Supplementary Fig. 5d). All higher DM IPHs exhibited responsiveness to temperature in a UCST fashion and HDO in a dose-dependent and recoverable manner. The critical temperature and the critical HDO concentration (for LLPS disruption) increased with increasing DM (Fig. 2d, e and Supplementary Fig. 5e, f).

Next, we investigated the effect of MW of dextran backbone on LLPS at fixed mass concentration, namely, fixed concentration of 'stickers'. The valency of IPH refers to the total number of oligopeptide 'stickers' on a single macromolecule, which is affected by MW of dextran backbone. The effect of branching[45] could be neglected, as the degree of branching (DB) of backbone dextran with different MW was confirmed to be comparable[46] (Supplementary Fig. 1a). The valency of $IPH_{Dex-6k}$, $IPH_{Dex-40k}$, $IPH_{Dex-550k}$ and $IPH_{Dex-2000k}$ was about 15, 94, 1300 and 4605, respectively. Higher MW IPHs could exhibit prominent LLPS under physiological conditions (Fig. 2f and Supplementary Fig. 6a, b), in contrast with the dispersed phase behavior of $IPH_{Dex-6k}$, thereby underscoring significance of valency for initiating LLPS. This result is consistent with an observation of LLPS of IDP in living cells that higher valency allows phase separation at a lower fractional saturation of 'stickers'[44]. Enzymatic cleavage of dextran backbone could undermine LLPS in a time/dose-dependent and highly efficacious fashion, further supporting the significance of multivalency for maintaining LLPS (Supplementary Fig. 6g). A plausible explanation is that LLPS is formed *via* a two-step nucleation-growth pathway. At a physiologically relevant mass concentrations (0.697 mg/mL[32,36]), IPH is under dilute polymer solution regime where the concentration is too low for polymer chains to interpenetrate[47] (see Supplementary Methods). Initially, the intramolecular interactions from conjugated oligopeptides dominate the seeding process, owing to the spatial proximity *via* backbone linkage, by providing the driving force for the formation of nuclei. This step is followed by maturation and micron-scale droplets formation over time which involves intermolecular interactions. Therefore, only moderate-MW IPHs ($IPH_{Dex-40k}$ and $IPH_{Dex-550k}$) phase-separated into large and numerous droplets. For extremely small MW IPH ($IPH_{Dex-6k}$), nucleation failed owing to insufficient intramolecular interactions. For IPH with extremely large MW ($IPH_{Dex-2000k}$), only small droplets were formed because the growth step was hampered by the longer intermolecular distance. With higher MW, IPH was more prone to phase-separate under higher ionic strength (Fig. 2g and Supplementary Fig. 6c) and exhibited robust LLPS under a wider pH range (Fig. 2h and Supplementary Fig. 6d). All phase separated IPHs exhibited responsiveness to temperature in a UCST fashion and HDO in a dose-dependent manner, with higher-MW IPHs showing lower sensitivity under the conditions examined (Fig. 2i, j and Supplementary Fig. 6e, f). Notably, IPH* (IPH$_{Dex-40k}$), designed through structural mimicking of FUS, could undergo prominent LLPS under physiological conditions (in contrast with the dispersed state of IPH$_{Dex-6k}$), withstanding a larger range of ionic strength and showing more responsiveness to temperature and HDO (in comparison to MW IPH$_{Dex-2000k}$). IPHs of medium-MW (IPH$_{Dex-40k}$ and IPH$_{Dex-550k}$) exhibited similar phase separation behavior, suggesting that under conditions where fractional saturation of stickers is fixed, an optimal window of valency exists.

We further investigated the effect of tyrosine/arginine (Y/R) ratio on the LLPS behavior. IPHs with varying Y/R ratios were able to undergo LLPS, despite differing in the extent (Fig. 2k and Supplementary Fig. 7a, b). As indicated by the turbidity measurement, IPHs phase separated to a less extent at Y/R=0, 0.268 and $\infty$, and more prominently at the intermediate Y/R values of 1.03 and 3.25 (Fig. 2k and Supplementary Fig. 7a, b). For IPH with Y/R=$\infty$, only CLCDP oligopeptides are present. The result indicates that the RAC-harboring 'stickers' CLCDP alone are sufficient to drive LLPS and recapitulate the formation of micron-sized liquid compartments by FUS, in contrast to the irregular solid assemblies formed by Aβ peptide-inspired conjugates (Dex-CAβACP) (Supplementary Fig. 2d, 2e, 3e and 7g). The addition of CRP, the charge-containing oligopeptides, to IPH was found to increase the responsiveness of LLPS to changes in ionic strength (Fig. 2l and Supplementary Fig. 7c). IPHs at all Y/R ratios showed UCST phase behavior (Fig. 2n and Supplementary Fig. 7e) and HDO responsiveness (Fig. 2o and Supplementary Fig. 7f), reflecting the labile and dynamic nature of phase-separated liquid droplets. Note that the extent of LLPS of IPH* (Y/R=1.03) was most sensitive to changes in ionic strength (Fig. 2l and Supplementary Fig. 7c), temperature (Fig. 2n and Supplementary Fig. 7e) and HDO (Fig. 2o and Supplementary Fig. 7f), while remaining robust over a broader range of pH (Fig. 2m and Supplementary Fig. 7d). The stimuli-responsive behavior of IPH system containing dual peptides should be beneficial for imparting function, such as the reversible recruitment and release of biomolecules.

**IPH\* droplets as artificial MOs**

The LLPS of IPH\* implies potential of displaying functionalities of natural MOs. MOs act as subcellular condensates that enrich various biomolecules including RNA and proteins[48,49] and host biochemical reactions[2]. Thus, we investigated whether IPH\* droplets could function as artificial MOs, in terms of preferential recruitment of compositional macromolecules and compartmentalized reaction enhancement. Model RNA and protein molecules, namely, polyuridylic acid (polyU) and green fluorescent protein (GFP), were both recruited and highly enriched within AMOs with 716 ($\pm$132) and 102 ($\pm$29.8) folds of enrichment, respectively (Fig. 3a, b and Supplementary Table 1). The charge-charge attractions between arginine residues from IPH\* and phosphate groups and the aromatic interactions between tyrosine residues of IPH\* and uracil of polyU could explain the recruitment of RNA. Similarly, the favorable enrichment of GFP could be attributed to non-specific charge-charge and aromatic interactions between IPH\* and the protein. Moreover, AMOs demonstrate reversible release and recruitment of GFP and polyU in response to temperature change in physiologically relevant range (Fig. 3a, b). Addition of polyU was found to promote LLPS of IPH\* when the amount reached a stoichiometric ratio or above (Supplementary Fig. 8b, c), presumably by reinforcing the network interactions within droplets. Note that sphericity was maintained for all the RNA containing condensates, implying the preservation of the liquid-like property. Other model RNAs (poly A and tRNA) tested also showed similar modulation of IPH\*'s LLPS behavior (Supplementary Fig. 8d, e).

Next, we demonstrated the possibility to carry out compartmentalized catalysis using horseradish peroxidase (HRP)-catalyzed decomposition of hydrogen peroxide as a model reaction with the fluorescent resorufin as a reporting molecule (Fig. 3c). HRP was enriched in IPH* droplets by 246 ($\pm$ 65.5) folds, thereby confining the reaction inside the liquid compartment. This was confirmed by real-time confocal imaging, which demonstrated gradual change of red fluorescence in the droplet interior (0–480s) (Fig. 3d). Moreover, the increase in fluorescence signal was much faster in the condensed phase than in the dispersed phase (Fig. 3e), supporting that the local reaction rate was about 15 times faster in the droplet. Combined reactions (in both condensed and dispersed phase) were quantified by time-dependent absorbance change of resorufin in bulk solution. The overall reaction rate in the presence of LLPS was also higher than that in the absence of LLPS ($v_0 = 51 \pm 2.2 \text{ nM} \cdot \text{s}^{-1}$ versus $v_0 = 22 \pm 6.2 \text{ nM} \cdot \text{s}^{-1}$). Note that the enhancement of overall reaction rate is not that significant as local reaction rate, owing to the limited volume fraction of condensed phase ($\phi_{\text{con}} = 0.32 \pm 0.026 \text{ \%}$, Supplementary Table 1). These results show AMOs formed by IPH* have the potential not only to mimic biophysical properties but functions of MOs in providing a dynamic and hierarchical organization of biomolecules.

## Conclusions

By describing an IDP FUS with a "stickers-and-spacers" model, we designed a minimalistic representation of the protein. The IPH, containing short peptides derived from RACs and arginine-rich sequences grafted onto a flexible polymer backbone, exhibited LLPS behavior reminiscent of the formation of natural MOs. Systematic variation of DW, MW, Y/R ratio further reviewed the molecular determinants of LLPS of IPHs, and agreement was found with FUS. The droplets formed by IPH acted as artificial MOs, enabling recruitment and enrichment of model RNAs and proteins, and providing liquid compartments for localizing and enhancing an enzymatic reaction. We believe that IPHs afford simple yet useful model systems for elucidating molecular interactions for the assembly of MOs. As a new type of biomaterials, IPHs create new possibilities for the dynamic delivery of proteins, nucleic acids, as well as in situ biochemical reactions.

## Acknowledgements

Financial support was provided by the Hong Kong Research Grant Council (GRF 16102520 and GRF 16103517) and the Hong Kong PhD Fellowship Scheme (for Jianhui Liu). We thank Dr. Rong Ni and Dr. Laurence Lau for useful discussions during this research.

## Author Contributions

J.L. and Y.C. conceived the concept. J.L., F.Z. and Y.C. designed the experiments. J.L., F.Z. performed experiments, data analysis and interpretation, whilst X.D. and J.H. provided guidance and assistance to the optical tweezer experiments. J.L., F.Z. and Y.C. wrote the manuscript. Y.C. supervised the research. All authors contributed to the discussion in preparation of the manuscript.

# Figures

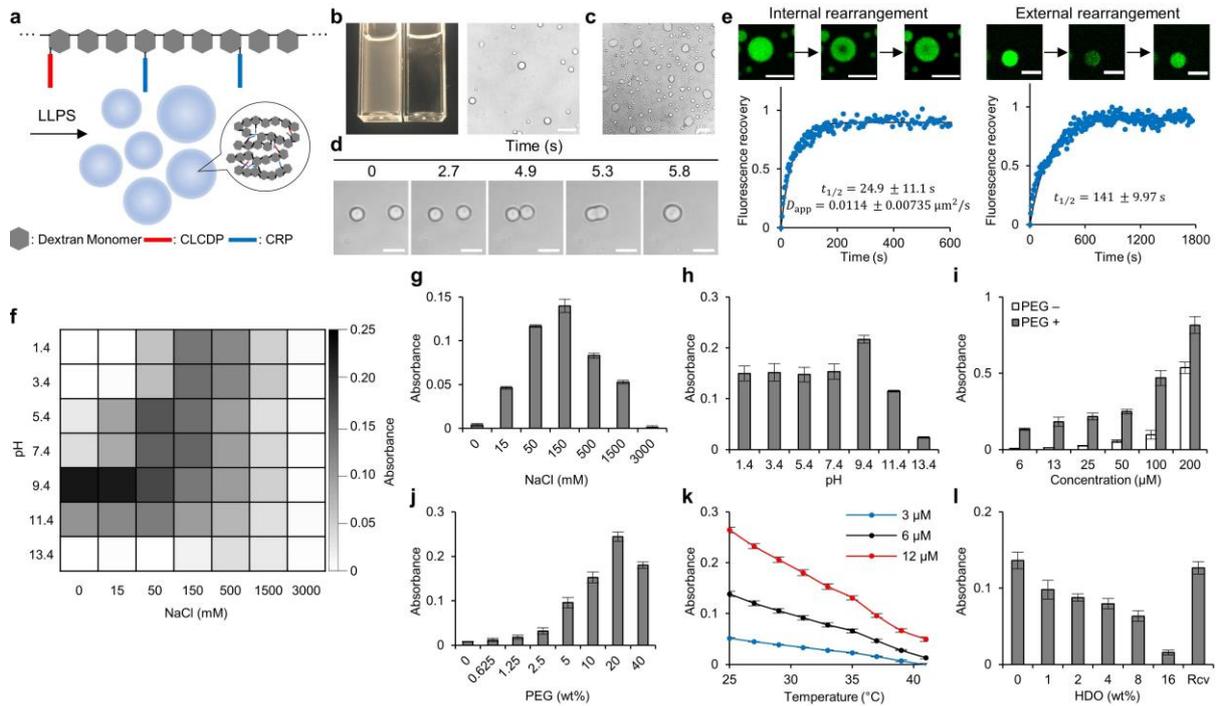

**Fig. 1. LLPS of IPH* generates compartmentalized micron-sized droplets. a**, Schematic illustrating LLPS of multivalent IPH. **b**, Micro-compartmentalization formation indicated by turbidity change (left, left cuvette: IPH* solution; right cuvette: cognate buffer for comparison) and micron-scale droplets formation (right). Scale bar, 20 μm. **c-e**, Liquid-like nature of compartmentalized droplets indicated by wetting phenomenon (glass without passivation, scale bar, 20 μm, **c**), fusion event (induced by optical tweezer, scale bars, 5 μm, **d**) and FRAP (indication of both internal and external molecular rearrangement, scale bars, 10 μm, **e**). **f-h**, LLPS is dually responsive to ionic strength and pH. **i**, **j**, IPH phase separates in a concentration-dependent (**i**) and crowding condition-dependent (**j**) manner. **k**, **l**, LLPS is sensitive to temperature (**k**) and HDO (**l**) stimuli.

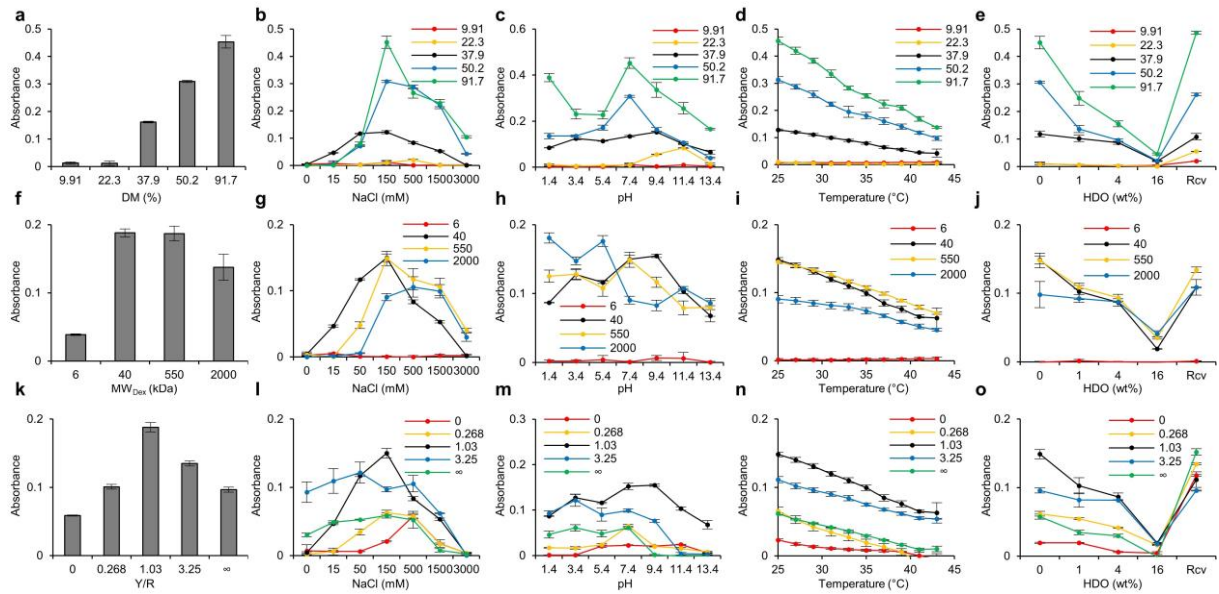

**Fig. 2. Molecular structural features modulate LLPS of IPH. a-e**, Modulation of DM contributes to modification of LLPS propensity (**a**), responsiveness to ionic strength (**b**), pH (**c**), temperature (**d**) and HDO disruption (**e**). **f-j**, Backbone MW of IPH affect LLPS propensity (**f**), responsiveness to ionic strength (**g**), pH (**h**), temperature (**i**) and HDO disruption (**j**). **k-o**, Modulation of Y/R contributes to modification of LLPS propensity (**k**), responsiveness to ionic strength (**l**), pH (**m**), temperature (**n**) and HDO disruption (**o**).

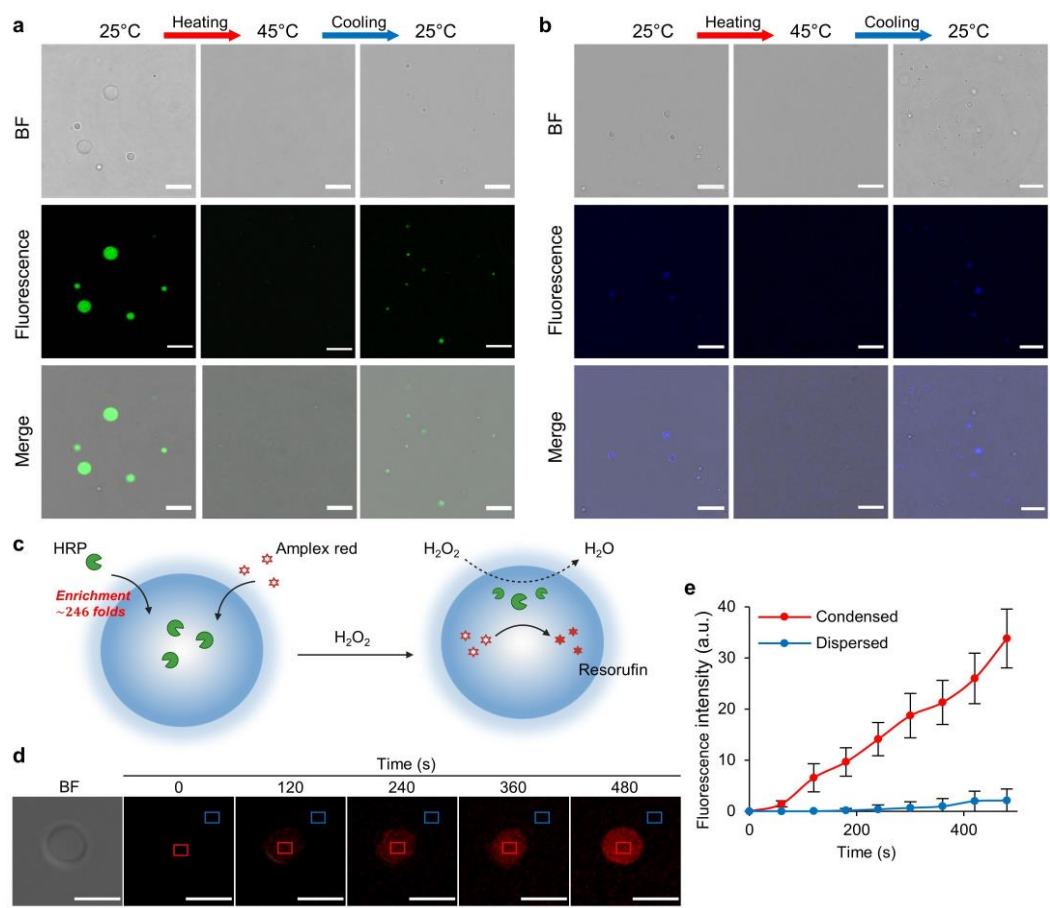

**Fig. 3. Functioning of AMOs. a**, **b**, Dynamic recruitment and release of protein (GFP at 60 μg/mL loading, **a**) and RNA (polyU incorporated at N/P=0.01, methylene blue added at 25 μM, **b**) by AMOs. IPH* droplets release protein/RNA upon heating to 45 °C concomitant with droplets dissolution, whereas recruiting protein/RNA upon re-cooling to 25 °C concomitant with droplets formation. **c**, Schematic illustrating enzymatic reaction in the IPH* droplet. Diffusion of the hydrogen peroxide into the IPH* droplet containing HRP initiates oxidation of the Amplex Red and its conversion into fluorescent resorufin. **d**, Time-lapse fluorescent confocal microscopy showing production of resorufin within the IPH* droplet. **e**, Average fluorescence intensity change over time in the interior (condensed phase) and exterior (dispersed phase) of the IPH* droplet (n=6). Scale bars,10 μm.

# Supplementary Information

**Minimalist design of polymer-oligopeptide hybrid as intrinsically disordered protein-mimicking scaffold for artificial membraneless organelle**

Jianhui Liu [1], Fariza Zhorabek [1], Xin Dai [2], Jinqing Huang [2] and Ying Chau [1,*]

Affiliations:

1. Department of Chemical and Biological Engineering, the Hong Kong University of Science and Technology, Clear Water Bay, Kowloon, Hong Kong.

2. Department of Chemistry, the Hong Kong University of Science and Technology, Clear Water Bay, Kowloon, Hong Kong.

# Materials and Methods

## Materials

Dextran from *Leuconostoc* spp. (approximate MW: 6, 40, 550 and 2000 kDa), PEG 8000 (as the crowding agent) and Dextranase from *Chaetomium erraticum* were purchased from Sigma-Aldrich (USA). Model RNAs, including polyuridylic acid potassium salt (polyU, *ca*. 2.47-3.09 kb), polyadenylic acid potassium salt (polyA, *ca*. 0.6-4 kb), tRNA from baker's yeast (*ca*. 0.071-0.084 kb), were purchased from Sigma-Aldrich (USA). Horseradish peroxidase (HRP) were purchased from Sigma-Aldrich (USA). THE RNA storage solution, RNaseZap RNase decontamination wipes, RNAsecure RNase inactivation reagent (for the treatment of solution) and nuclease-free water were purchased from Thermo Fisher Scientific (USA). Amplex Red (10-acetyl-3,7-dihydroxyphenoxazine) and hydrogen peroxide (3% solution) were purchased from Thermo Fisher Scientific (USA). Quant-iT RiboGreen RNA Assay Kit was purchased from Thermo Fisher Scientific (USA). FITC-PEG1000-SH (purity>95%) was purchased from Shanghai Ponsure Biotech (China). All the other chemicals were purchased from Sigma-Aldrich (USA) unless otherwise specified. Peptides were synthesized *via* solid-phase peptide synthesis method and purified (purity>95%) with high-performance liquid chromatography (HPLC) by Sangon Biotech (China). The identities of peptides were double confirmed by mass spectroscopy (Supplementary Fig. 2e) using mass spectrometer (Bruker Daltonics UltrafleXtreme MALDI TOF/TOF Mass-spectrometer, USA).

## Methods

**Quantification of DB of dextran**

DBs of dextran were quantified from $^1$H NMR (on 500 MHz Varian NMR spectrometer, USA, dissolved at 10 mg/mL in deuterium oxide) by comparing integrations of branching peaks with main chain peak[1].

**Synthesis of Dex-VS with desired DM**

Dex-VS was synthesized following an one-step Michael-addition click chemistry route[2]. 300 mg dextran was dissolved in 30 mL ultrapure H$_2$O (UPW). Under rigorous stirring, 120 μL NaOH (5M) was added to adjust the pH to alkaline condition ([OH$^-$]=20 mM, pH=12.3). 669 μL DVS (1.2 times stoichiometrically excess to hydroxyl on dextran) was added quickly into solution. After certain reaction time, 126 μL HCl (5M) was quickly added (1.05 times stoichiometrically excess to NaOH) to adjust the pH to mild acidic condition (pH=3.0) to terminate the reaction, followed by dialysis against 3 L UPW using dialysis tubing (Spectra/Por Dialysis Membrane, Spectrum Labs, USA) for 5 days with 10 times UPW replacement. The MW cut-offs (MWCOs) for 6 kDa dextran and the others are 1 kDa and 3.5 kDa, respectively. Purified Dex-VS aqueous solution was lyophilized and stored in −20 °C for further usage.

DM, defined as number of conjugated VS groups divided by number of repeating units[2], was quantified from $^1$H NMR by comparing integration of VS $^1$H peaks with integration of dextran $^1$H1 peak.

**Determination of DM of IPH to mimic IDPs**

Within the LCDs of IDPs, the patterning of RACs was quantified by average spacing of RAC ($\bar{L}$):

$$\bar{L} = \frac{X}{n} - 1$$

To mimic the patterning of RACs, DM of IPH was designed as:

$$DM = \frac{1}{\bar{L} + 1}$$

where

$X$: Number of residues of LCD

$n$: Number of RACs harbored by LCD

**Synthesis of IPH**

The chemistry of synthesis is thiol-Michael addition reaction between thiol groups (from cysteine) and VS groups. NaCl (3 M) solution in 25 mL flask was pretreated for 30 min under $N_2$ flow, followed by peptides dissolution and 10 min $N_2$ flow treatment for potential disulfide bond reduction. Subsequently, Dex-VS was added and dissolved under rigorous stirring, followed by acidity adjustment to pH=6.0 for reaction initiation with minute amount of NaOH (5 M). $N_2$ flow was maintained for another 10 min, then the flask was sealed under stirring for 24 h-reaction. The materials were then transferred to dialysis tubing (MWCO=3.5 kDa) against 3 L UPW for 5 days with 10 times UPW replacement. After dialysis, the IPH aqueous solution was diluted for 4 times using UPW, followed by lyophilization and −20 °C storage for further usage. The structural identity was characterized with $^1$H NMR (dissolved at 10 mg/mL in deuterium oxide hereafter unless otherwise specified). DMs of peptides were quantified by comparing integrations of peptide characteristic peaks with integration of dextran $^1$H1 peak.

For a specific batch synthesis, 15 mg Dex-VS was dissolved at 2 mg/mL. The total molarity of peptides was 1.5 times stoichiometrically excess to [VS] of Dex-VS. The Y/R of IPH was controlled by tuning the molar feed ratio of reactant peptides ($FR_\text{p}$), namely,

$$FR_\text{p} = \frac{n_\text{CLCDP}}{n_\text{CRP}}$$

Note that IPH* corresponds to the IPH that we are of the most interest, which mimics FUS protein in terms of triple aspects, namely, number of repeating units (by MW), motif patterning (by DM) and Y/R (by ratio of conjugated peptide), to the most extent. Dextran (40 kDa) was modified with VS groups at 37.9 % DM, followed by peptide conjugation at Y/R=1.03 (Supplementary Fig. 3a).

**Synthesis of IPH* with FITC labeling**

A two-step one-pot synthesis method was applied. 5 mL NaCl (3 M) aqueous solution in 10 mL flask and FITC-PEG1000-SH solution (in DMF at 5 mg/mL) was pretreated for 30 min under $N_2$ flow, followed by Dex-VS (10 mg) addition under rigorous stirring until complete dissolution (at 2 mg/mL). Subsequently, 90.6 μL FITC-PEG1000-SH solution (feeding fluorophore was 1.25 times stoichiometrically excess to IPH) was added very slowly (in 5 μL small aliquots), followed by pH adjustment to 6.0 using minute (around 0.15 μL) NaOH (5 M) and a 30-min reaction under rigorous stirring and $N_2$ flow. Peptides were then premixed, added and dissolved, followed by pH adjustment to 6.0 using minute (around 0.15 μL) NaOH (5 M) and a 10-min reaction under rigorous stirring and $N_2$ flow. The materials were then transferred to dialysis tubing (MWCO=12-14 kDa) against 3 L UPW for 5 days with 10 times UPW replacement. After dialysis, the IPH aqueous solution was diluted for 4 times using UPW, followed by lyophilization and −20 °C storage until further usage. The structural identity was characterized with $^1$H NMR and DMs of peptides were quantified by comparing integrations of peptide characteristic peaks with integration of dextran $^1$H1 peak. Dark condition was maintained for the whole synthesis, characterization and sample storage process. $^1$H NMR data was comparable with unlabeled IPH* (data not shown). The F/P (fluorophore/IPH) ratio was determined to be 0.461 by quantification of fluorescence intensity of fluorophores.

**LLPS formation from IPH**

For a prototypical preparation of IPH* physiological-mimicking solution, the stock solution (50 x) of IPH* was prepared by dissolving IPH* in UPW at 34.9 mg/mL (approximately 300 µM). IPH* stock solution was stored in −80 °C as single-use aliquots. For sample preparation, IPH* stock solution was heated to 70 °C and incubated for 2 min in sealed centrifuge tubes, followed by dilution in 25 °C intracellular physiology-mimicking buffer (IPM buffer, as the default buffer unless otherwise specified, containing 150 mM NaCl, 10 mM HEPES, 10 wt% PEG 8000, pH=7.4) at 0.697 mg/mL (approximately 6 µM), brief vortex mixing and 15-min incubation at 25 °C before characterization. For reference, intracellular concentration of FUS protein in HeLa cells[3] and Sf9 cells[4] ranges from 1 µM to 8 µM (0.0534 mg/mL to 0.427 mg/mL).

For the other IPHs with different structure, the stock (50x) concentration was used differently, whilst the rest of protocols are identical. For an IPH with different DM and Y/R, the stock (50 x) was prepared at 300 µM molarity. For an IPH with different MW, the stock (50 x) was prepared at 34.9 mg/mL mass concentration. Note that at physiology-mimicking mass concentrations (0.697 mg/mL), IPH with all MWs are considered as under dilute polymer solution regime, as critical concentration $c^*$ of dextran can be determined based on intrinsic viscosity[5] $[\eta]$ via $c_\eta^* = \frac{2.5}{\eta}$ relationship[6]. For dextran used in this paper, 0.697 mg/mL $\ll$ $c_{\eta,\text{min}}^* = c_{\eta,\text{Dex}-2000k}^* = 41$ mg/mL.

For LLPS formation under different solvent conditions (salt, pH and crowding agent), the investigated parameter was modulated while the rest are fixed at physiology-mimicking conditions.

**Turbidimetry assay**

Turbidimetry assays were performed on microplate reader (Varioskan LUX multimode, Thermo Fisher Scientific, Finland), wherein 40 µL solution was loaded to 384-well plate (polystyrene, F-bottom, clear, non-binding, Greiner Bio-One, Germany). Absorbance at 600 nm was measured as the quantification of turbidity. Each individual sample preparation was triplicated. Turbidity for each well is taken by averaging 3 times of measurement with blank corrected.

**HDO assay**

HDO disruption-recovery assay was conducted for the characterization of lability of IPH assemblages. For HDO disruption assay, HDO was introduced as the last component to the IPH solution. For HDO recovery assay, IPH, disrupted with 16 *wt*% HDO, was transferred to the dialysis tubing (in 0.5-1 kDa MWCO tubing) against 3 L UPW for 5 days with 10 times UPW replacement. After dialysis, the IPH aqueous solution was diluted for 4 times using UPW, followed by lyophilization and sample dissolution in buffer (containing 150 mM NaCl, 10 mM HEPES, pH=7.4). We assume complete retention of PEG 8000 and IPH during the dialysis process owing to large size difference from MWCO.

**Microscopic imaging**

Imaging under room temperature was performed on confocal microscope (Leica TCS SP8, Germany) using 100x oil objective lens by loading 10 µL of sample solution to a confocal dish. Imaging under heated conditions was performed on Nikon C2 confocal microscope with 60x lens using 500 µL solution loaded onto a sealed confocal dish. All dishes were passivated with Pluronic F-127 (10 wt%) for 1 hour, prior to the experiment, to lower material adhesion unless otherwise specified.

**Fluorescence recovery after photobleaching (FRAP)**

FRAP experiments were performed on the laser scanning confocal microscope (Nikon C2, Japan) using NIS-Elements software. Photobleaching was performed using 488nm Argon laser with 50% intensity. For photobleaching within droplet, circular bleached spots with 2 μm diameter within ∼6–8μm droplets were used for measurements, while for full droplet bleaching, entire size of the droplet was bleached. Fluorescence intensity changes over time following bleaching were recorded for 3 different regions of interest (ROIs, including bleached spot, reference, and background) at 1% laser intensity until fluorescent signals in bleached region recovered and reached an equilibrium state.

Data analysis was performed following reported protocol[7]. Values were first background subtracted and corrected for any photofading during image acquisition:

$$I(t) = \frac{F(t)R_\text{i}}{F_\text{i}R(t)}$$

where

$F_i$:  initial intensity of bleached ROI

$F(t)$:  intensity of bleached ROI at time $t$

$R_i$:  initial intensity of reference ROI

$R(t)$:  intensity of reference ROI at time $t$

$I(t)$:  corrected intensity of bleached ROI.

Full-scale normalization was performed on relative fluorescence intensity data, assigning minimum bleached intensity to 0 and pre-bleach intensity to 1.

$$I_{\text{norm}}(t) = \frac{I(t) - I_0}{I_i - I_0}$$

$I_0$:  corrected intensity of ROI immediately after bleaching

$I_i$:  corrected initial (pre-bleach) intensity of ROI

Normalized data was fitted to single exponential equation using MATLAB:

$$I_{\text{norm}}(t) = A(1 - e^{-\frac{t}{\tau}})$$

where $\tau$ is the recovery time constant and A is constants, corresponding to amplitude of recovery. Half-time recovery timescale was further obtained using $\tau$ value:

$$t_{1/2} = \tau \cdot \ln 2$$

The apparent diffusion coefficient ($D_{\text{app}}$) was obtained using Axelrod model equation for 2D-diffusion[8]:

$$D_{\text{app}} = \frac{\gamma_D w^2}{4 t_{1/2}}$$

With $w$ being the radius of bleached region, and $\gamma_D$ laser-beam shaped-dependent constant value, which equals 0.88 for uniform circular beam.

**Fusion from optical tweezer**

Controlled coalescence of suspended IPH* droplets were conducted using dual optical trap (m-Trap$^{TM}$, LUMICKS). Droplet sample was prepared at 6 µM of IPH and incubated for 15 mins under room temperature. Samples were loaded into 1.0 x 1.0 cm chamber (Gene Frame) fixed on a glass slide treated with 10% Pluronic-F. IPH* droplets were trapped with a 1064 nm laser at 20 % laser power, that was found to be optimal for stable trapping of our droplets. Trapping of the suspended droplets was achieved owing to a difference in refractive index between the IPH* droplets and buffer. After trapping, one droplet remained at stationary Trap-2 while another droplet on Trap-1 were set to move at a constant velocity of 0.13 µm/s until surfaces of both droplets were brought into contact. The Trap-1 movement was stopped once trapped droplets fused and relaxed back to a spherical share. Time series were recorded for each fusion event.

**Quantification of RNA concentration of stock**

The concentration of RNA stock solution was determined on NanoVue Plus spectrophotometer (Biochrom, USA). Measurement was triplicated while the average value was taken.

**Enrichment of RNA/protein from AMO**

Droplets from IPH were pre-formed and incubated for 15 min as aforementioned in 'LLPS formation from IPH' section. The polyU was first selected as the model RNA owing to its random coil structure[9]. For RNA enrichment, RNAs in the stock solution (*c.a.* 50x concentrated, in RNA storage solution) was pipetted into IPH solutions, followed by gentle pipette mixing and 15-min incubation. Methylene blue (MB), the staining reagent that binds to nucleic acids[10], was then incorporated lastly at 10 μM concentration and incubated for 15 min before imaging, given that MB *per se* shows no evident preferential partitioning in IPH* solution (Supplementary Fig. 8a). MB-stained RNA was visualized using diode laser line at 638 nm. Nuclease-free condition was maintained for the whole process of RNA-related experiments following technical bulletin provided by Thermo Fisher Scientific (USA). GFP was selected as the model protein owing to the negative charge at physiological pH and innate fluorescent property. For protein enrichment, GFP in the stock solution (*c.a.* 50x concentrated) was pipetted into IPH droplet solutions, followed by gentle pipette mixing and 15-min incubation before imaging. Samples were excited by a 488-nm laser to visualize partitioning behavior of GFP.

The quantification of enrichment of cargoes was performed based on the volume fraction determination of droplets. IPH* droplets were pre-formed in IPM buffer followed by loading cargoes at desired concentration, bringing the IPH concentration to 6 μM at a final volume of 200 μL. Droplets with loaded cargoes were incubated for 15 mins to ensure equilibrium. Condensed (droplet) phase was separated from the dispersed phase *via* centrifugation (21,000 x *g*, fixed 25 °C, 15 min) and concentration of cargoes in the dispersed phase ($[Cargo]_{\text{dis}}$) was determined by comparing with established standard curves (with blank corrected). The volume fractions of the condensed phase $\phi_{\text{con}}$ were estimated using 3-D rendering (Supplementary Fig. 9) of the sample images following Ghosh et al[11]. We define dimensionless concentration ($DC$) in this partitioning system as:

$$DC = \frac{[Cargo]}{[Cargo]_{\text{ctrl}}}$$

'ctrl' means the control system without any partitioning (namely, without IPH*).

Based on mass balance, $DC$ of cargoes in the condensed phase, $DC_{\text{con}}$, was determined by:

$$DC_{\text{con}} = \frac{1 - (1 - \phi_{\text{con}})DC_{\text{dis}}}{\phi_{\text{con}}}$$

and $Enrichment$ was determined as:

$$Enrichment\ = \frac{DC_{\text{con}}}{DC_{\text{dis}}}$$

The experimental data of cargoes enriched, namely, GFP, polyU and HRP, were summarized as Supplementary Table 1.

**Catalysis of reaction from HRP**

To investigate the possibility to carry out biochemical reactions inside droplets, enzymatic activity of HRP was evaluated. Amplex Red was selected as a probe, which undergoes HRP mediated oxidation in the presence of hydrogen peroxide ($H_2O_2$) substrate, forming resorufin with fluorescence generated. Reaction was performed by first loading 100 ng/ml HRP and 50 μM Amplex Red into preformed IPH* droplets in IPM buffer, followed by 15 mins incubation for equilibrium of the loaded components. Sample was then deposited onto a confocal dish, and area to monitor the reaction was located under the microscope. To initiate the reaction, 1 μL of 500 μM $H_2O_2$ was added into the confocal dish without moving the sample and time lapse images were acquired immediately afterwards.

Imaging was performed on Nikon C2 scanning confocal microscope under excitation by 561 nm laser (0.35% power) using a 60x oil immersion lens. Time series were recorded with an interval of 60 s right after addition of hydrogen peroxide at 25 °C.

For the quantification of fluorescence intensity change inside and outside of the droplet, images were analysed by Fiji software. Fluorescence signal change was recorded in different droplets (n=6) and in different regions of the exterior (n=6), obtained from one single experiment.

To determine the global reaction rate of HRP catalysis, time-dependant absorbance change of resorufin at 560 nm was recorded on a 384 well-plate. Reaction rate was determined by:

$$v_0 = \frac{\Delta A}{\Delta t \cdot \varepsilon \cdot d}$$

$\frac{\Delta A}{\Delta t}$ :     change in absorbance at 560 nm over given time

$\varepsilon$ :     extinction coefficient of resorufin product (54000 $M^{-1}cm^{-1}$)

$d$ :     pathlength of light (0.42 cm)

# Supplementary Figures

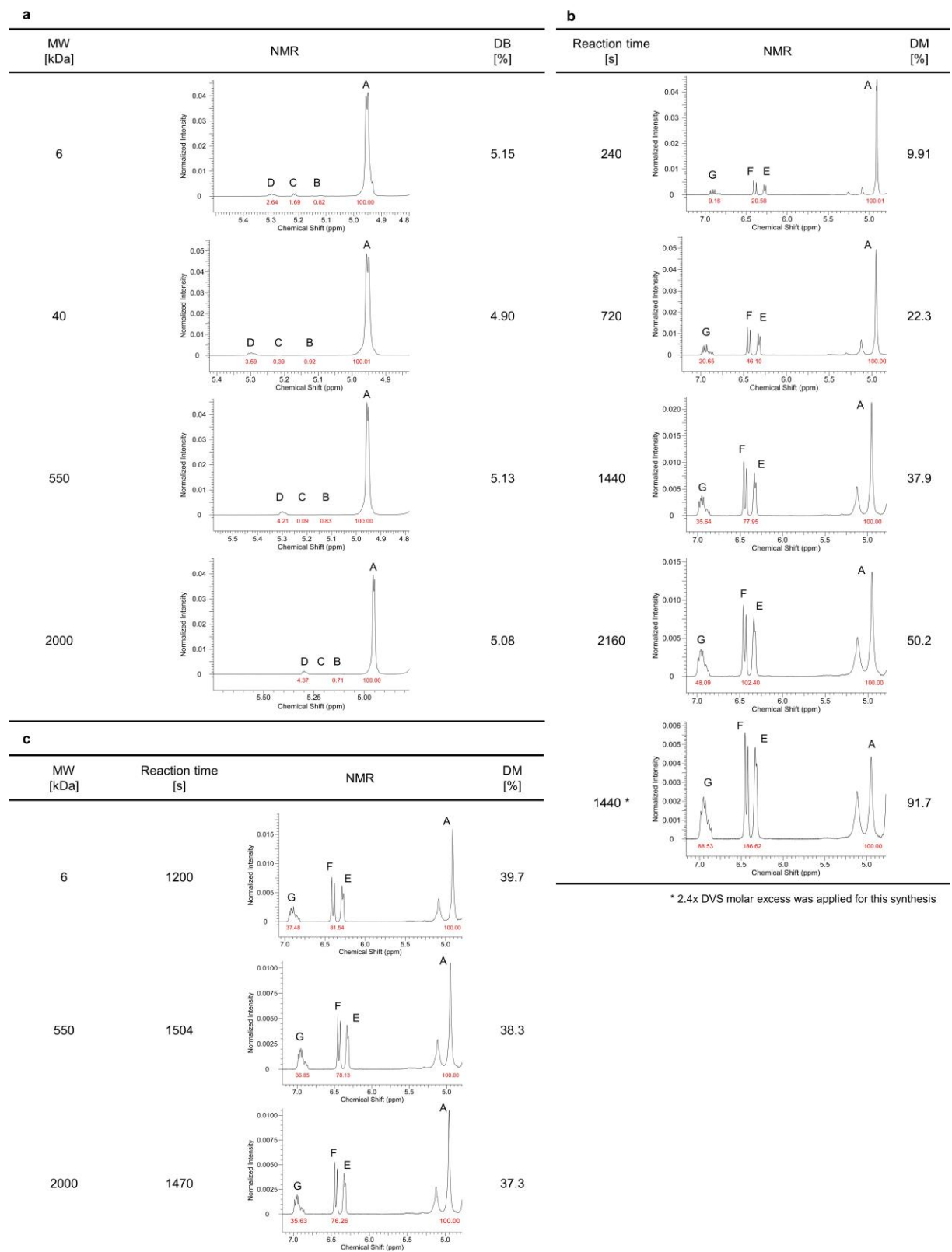

* 2.4x DVS molar excess was applied for this synthesis

**Supplementary Fig. 1. Characterization of dextran and Dex-VS. a**, Quantification of DB of dextran from $^1$H NMR data. DBs were quantified by comparing the sum of integrations of peak B, C and D (δ=5.11, 5.18 and 5.32 ppm, respectively) with peak A (δ=4.98 ppm)[1]. Peak A, B, C, and D corresponds to α-1,6 linkage (Main chain), α-1,2 linkage (1,2-Branching), 2,6-di-O-substituted glucopyranosyl unit and α-1,3 linkage (1,3-Branching), respectively. $DB = \frac{\int (B+C+D)}{\int A}$. The results indicate DB=5% ($r.e. \leq 3\%$) for dextran with different MWs examined. **b**, Quantification of DMs of Dex40-VS from NMR data. **c**, Quantification of DMs of Dex-VS with different MWs from NMR data. DMs were quantified by comparing the sum of integrations of peak E, F and G (δ = 6.3, 6.4, and 6.9 ppm, respectively, corresponding to double bonds of free VS) with peak A (δ=4.98 ppm). $DM = \frac{\int (E+F+G)}{3 \int A}$.

**a**

| Protein | Sequence |
|---|---|
| FUS | MASNDYTQQATQSYGAYPTQPGQGYSQQSSQPYGQQSYSGYSQSTDTSGYGQSYSSYGQSQNTGYGTQSTPQGYGSTGGYGSSQSSQSSYGQQSSYPGYGQQPAPSSTSGSYGSSSQSSSYGQPQSGSYSQQPSYGGQQQSYGQQQSYNPPQGYGQQNQYNSSGGGGGGGGGGGGNYGQDQSSMSSGGGSGGGYGNQDQSGGGGSGGYGQQDRGGRGRGGSGGGGGGGGGYNRSSGGYEPRGRGGGRGGRGGMGGSDRGGFNKFGGPRDQGSRHDSEQDNSDNNTIFVQGLGENVTIESVADYFKQIGIIKTNKKTGQPMINLYTDRETGKLKGEATVSFDDPPSAKAAIDWFDGKEFSGNPIKVSFATRRADFNRGGGNGRGGRGRGGPMGRGGYGGGGSGGGRGGFPSGGGGGGGQQRAGDWKCPNPTCENMNFSWRNECNQCKAPKPDGPGGGPGGSHMGGNYGDDRRGGRGGYDRGGYRGRGGDRGGFRGGRGGGDRGGFGPGKMDSRGEHRQDRRERPY |

**b**

| Protein | Length of full-length protein [residues] | Length of LCD [residues] | LCM | $\bar{L}$ [residues] | DM of IPH [%] |
|---|---|---|---|---|---|
| FUS | 526 | 214 | [G/S][Y][G/S] | 9.70 | 9.35 |

**d**

| Name | Length [residues] | Sequence | MW [Da] |
|---|---|---|---|
| CLCDP | 9 | [CH$_3$CONH]-CGGSYSGYS-[CONH$_2$] | 921 |
| CRP | 6 | [CH$_3$CONH]-CGGRGG-[CONH$_2$] | 547 |
| CAβACP | 9 | [CH$_3$CONH]-CGGKLVFFA-[CONH$_2$] | 982 |

**c**

| Protein | Length of full-length protein [residues] | Tyr [residues] | Arg [residues] | Y/R |
|---|---|---|---|---|
| CELF4 | 485 | 14 | 17 | 0.826 |
| CSTF2 | 577 | 7 | 41 | 0.171 |
| CSTF2T | 616 | 7 | 40 | 0.175 |
| DAZ1 | 744 | 48 | 29 | 1.66 |
| DAZ2 | 438 | 46 | 13 | 3.54 |
| DAZ3 | 438 | 46 | 13 | 3.54 |
| DAZ4 | 390 | 38 | 13 | 2.92 |
| DAZAP1 | 406 | 14 | 18 | 0.778 |
| EWSR1 | 656 | 42 | 45 | 0.933 |
| FUS | 526 | 36 | 37 | 0.973 |
| hnRNPA0 | 305 | 15 | 16 | 0.938 |
| hnRNPA1a | 320 | 12 | 24 | 0.500 |
| hnRNPA1b | 372 | 19 | 25 | 0.760 |
| hnRNPA1L2 | 320 | 12 | 22 | 0.545 |
| hnRNPA2B1 | 353 | 22 | 25 | 0.880 |
| hnRNPA3 | 378 | 22 | 29 | 0.759 |
| hnRNPAB | 327 | 21 | 13 | 1.62 |
| hnRNPD | 355 | 21 | 14 | 1.50 |
| hnRNPDL | 420 | 24 | 33 | 0.727 |
| hnRNPH1 | 472 | 26 | 32 | 0.813 |
| hnRNPH2 | 449 | 25 | 31 | 0.806 |
| hnRNPH3 | 346 | 23 | 28 | 0.821 |
| PSPC1 | 523 | 6 | 52 | 0.115 |
| RBM14 | 669 | 47 | 41 | 1.15 |
| RBM33 | 1170 | 11 | 75 | 0.147 |
| SFPQ | 707 | 14 | 60 | 0.233 |
| TAF15 | 589 | 54 | 56 | 0.964 |
| TDP-43 | 414 | 8 | 20 | 0.400 |
| TIA1 | 386 | 17 | 13 | 1.308 |
| TIAL1 | 392 | 16 | 14 | 1.14 |

**e**

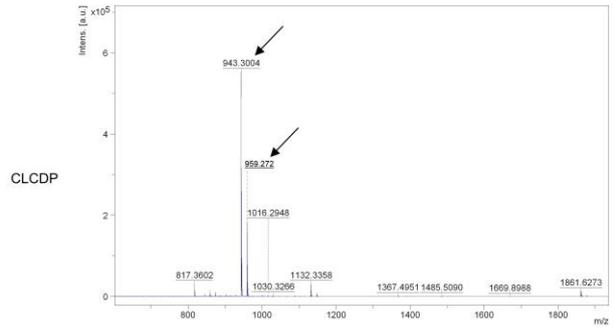

CLCDP

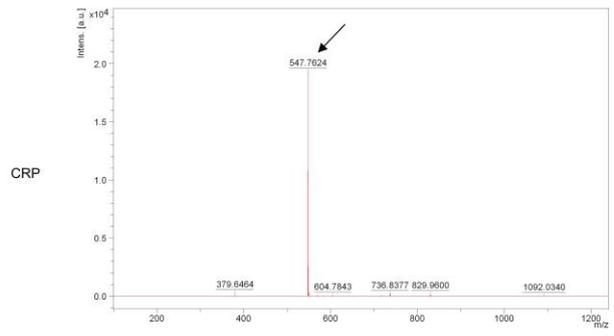

CRP

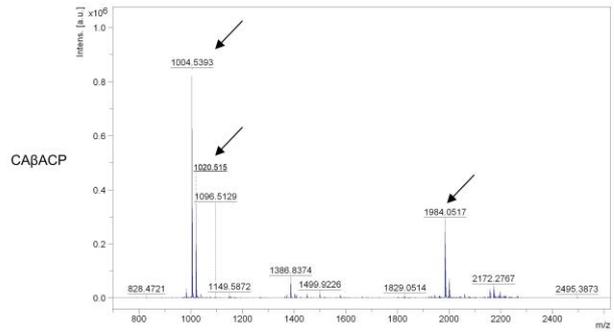

CAβACP

**Supplementary Fig. 2. Design and identity characterization of oligopeptides. a**, Sequences of FUS protein. LCDs and LCMs are underlined and highlighted, respectively. **b**, Determination of DM of peptides by corresponding LCM patterning. **c**, Investigation of Y/R of IDPs. The highest and the lowest Y/R values are highlighted in red and green color, respectively. **d**, Peptide sequences used in this research. N-terminal acetylation and C-terminal amidation were applied for eliminating associated charging effects. **e**, Identification confirmation of peptides from MALDI mass spectroscopy data. CLCDP: Theoretical MW= 921 Da; Measured MW= 921 Da ([M – H$^+$ + Na$^+$]=943 Da; [M – H$^+$ + K$^+$]=959 Da). CRP: Theoretical MW= 547 Da; Measured MW= 547 Da ([M + H$^+$]= 548 Da). CAβACP: Theoretical MW= 982 Da; Measured MW=982 Da ([M – H$^+$ + Na$^+$]=1004 Da; [M – H$^+$ + K$^+$]=1020 Da; [M$_2$ – H$^+$ + Na$^+$]=1984 Da). M$_2$ refers to dimers formed *via* disulfide bond linkages between cysteine residues.

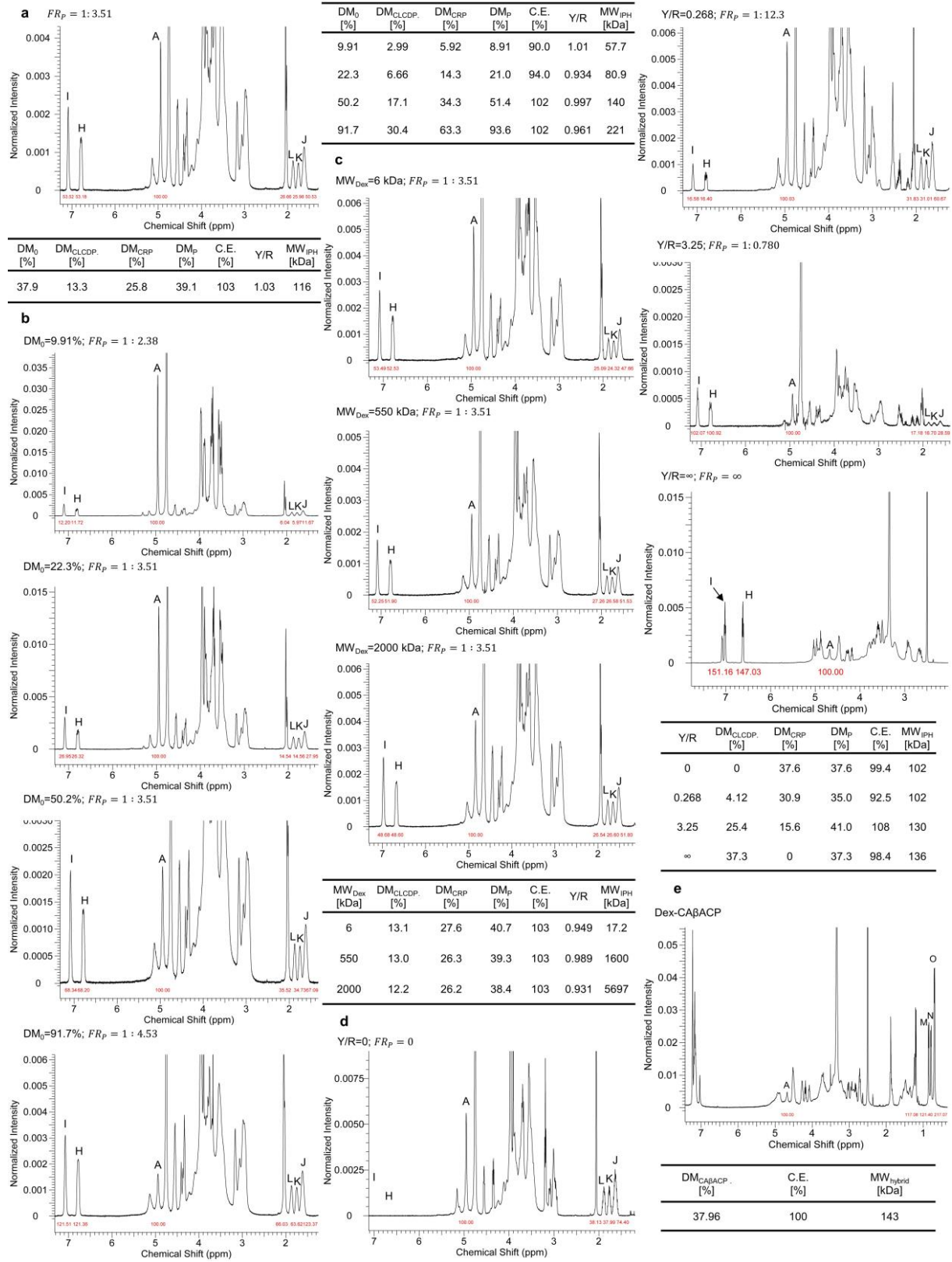

**Supplementary Fig. 3. Structural characterization of IPH.** $^1$H NMR of **a,** IPH*; **b**, IPH with different DMs; **c**, IPH with different MWs. **d**, IPH with different Y/R. Peak A (δ=4.98 ppm) corresponds to dextran main chain $^1$H1. DM of CLCDP was quantified by comparing integrations of peak H (δ=6.86ppm, corresponding to $^1$H3,5 of tyrosine) and peak I (δ=7.15ppm, corresponding to $^1$H2,6 of tyrosine) with peak A, namely, $DM_{\text{CLCDP}} = \frac{\int(H+I)}{8\int A}$. DM of CRP was quantified by comparing integrations of peak J (δ=1.70 ppm. corresponding to $^1$HC of arginine) and peak K, L (δ=1.79 ppm and 1.89 ppm, respectively, corresponding to $^1$HB of arginine) with peak A, namely, $DM_{\text{CRP}} = \frac{\int(J+K+L)}{8\int A}$. The total DM of peptides conjugated, $DM_P$, was further compared with the total DM of VS along the dextran backbone ($DM_0$) and the conjugating efficiency ($C.E. = \frac{DM_p}{DM_0}$) indicate satisfactory data consistency. Note that for Y/R=∞ sample of (**d**), DMSO-$d_6$ was used as the solvent, whilst peak A, H and I shift to δ= 4.69, 6.62 and 6.97 ppm, respectively. The (approximate) MW of IPH was quantified from addition of MW of dextran backbone, VS linkage segment and peptide conjugated. **e**, Dex-CAβACP for comparison. DM of CAβACP was quantified by comparing integrations of peak M, N and O (peak cluster, δ=0.7-0.9 ppm, corresponding to $^1$HG of valine and $^1$HD of leucine) with peak A, namely, $DM_{\text{CAβACP}} = \frac{\int(M+N+O)}{12\int A}$.

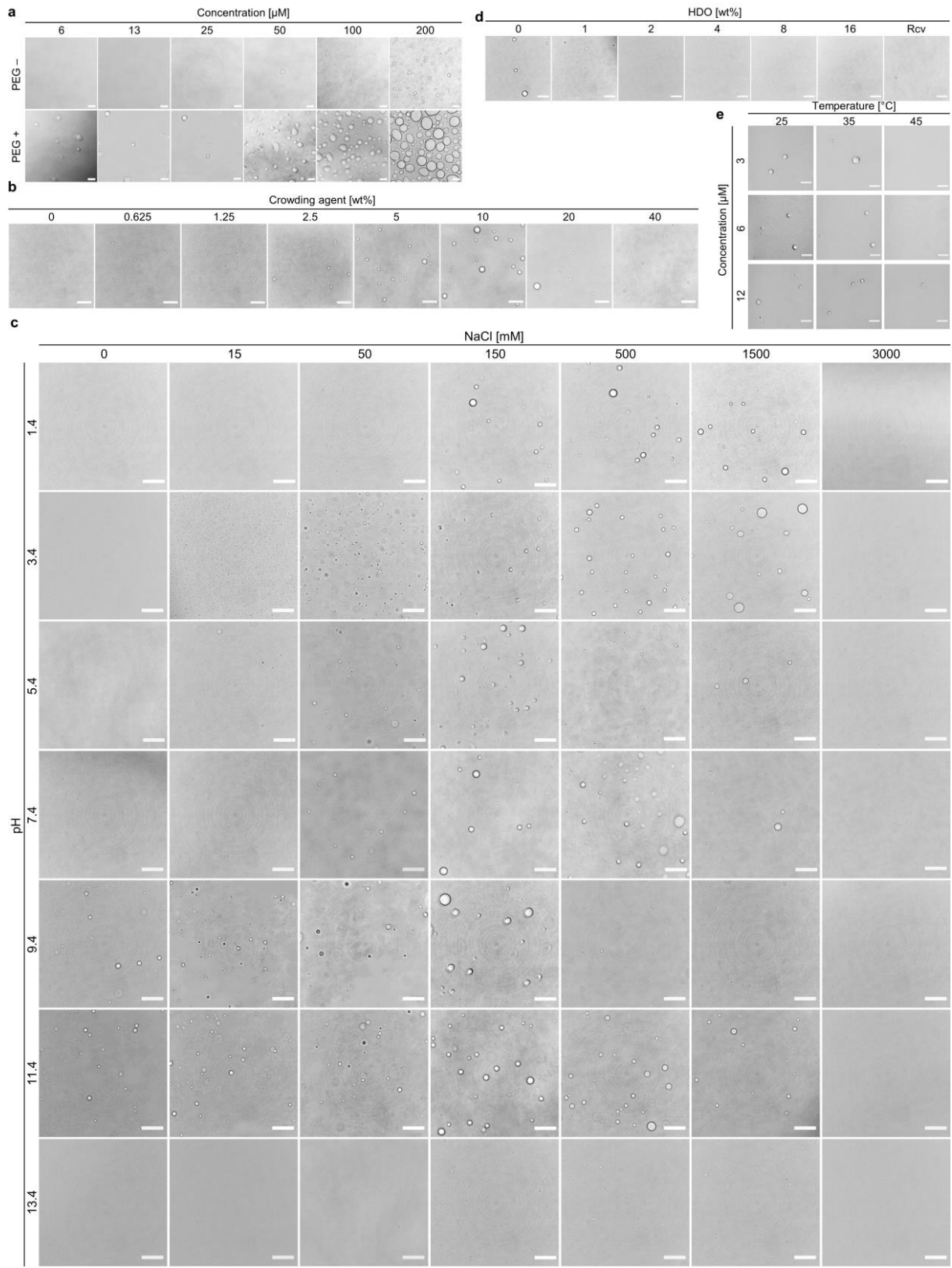

**Supplementary Fig. 4. Microscopic images of compartmentalized droplets from IPH\*. a**, Size and number of droplets are dependent of concentration and crowding agent conditions. **b**, Incorporation of crowding agent can facilitate (no higher than 20 wt%) or hinder LLPS formation (higher than 20 wt%). **c**, Droplet formation are dually responsive to ionic strength and pH. **d**, Droplet formation could be disrupted by incorporation of HDO, whilst removal of HDO could generate recovery of LLPS. **e**, Different concentrations of IPH\* solution exhibit LLPS with UCST phase behavior. Scale bars, 20 µm.

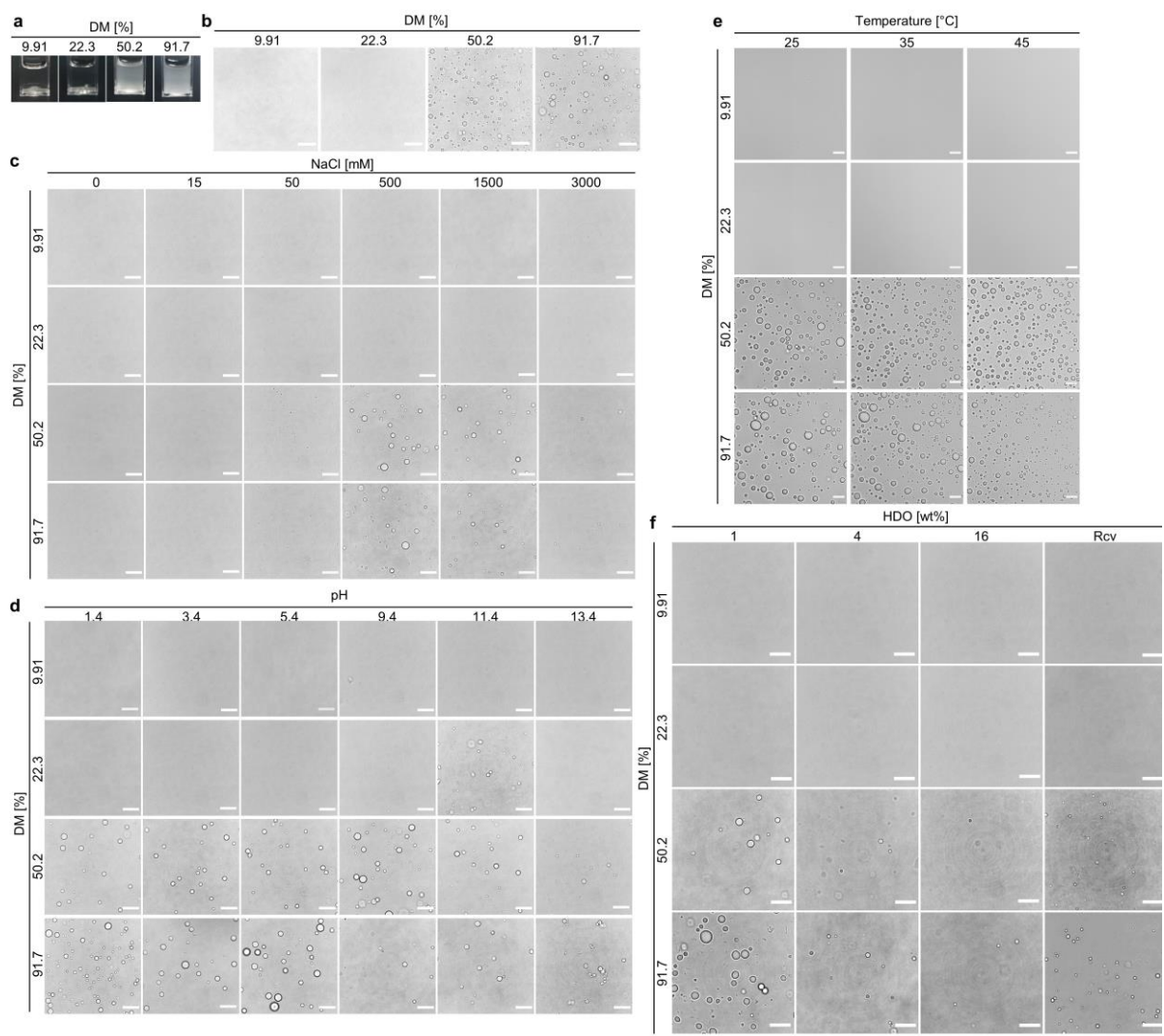

**Supplementary Fig. 5. Imaging LLPS of IPH with different DM. a**, Turbidity change indicates occurrence of LLPS. **b**, Microscale droplets formation. **c-e**, Responsiveness of LLPS to ionic strength (**c**), pH (**d**) and temperature (**e**). **f**, Reversibility of LLPS to HDO. Scale bars, 20 µm.

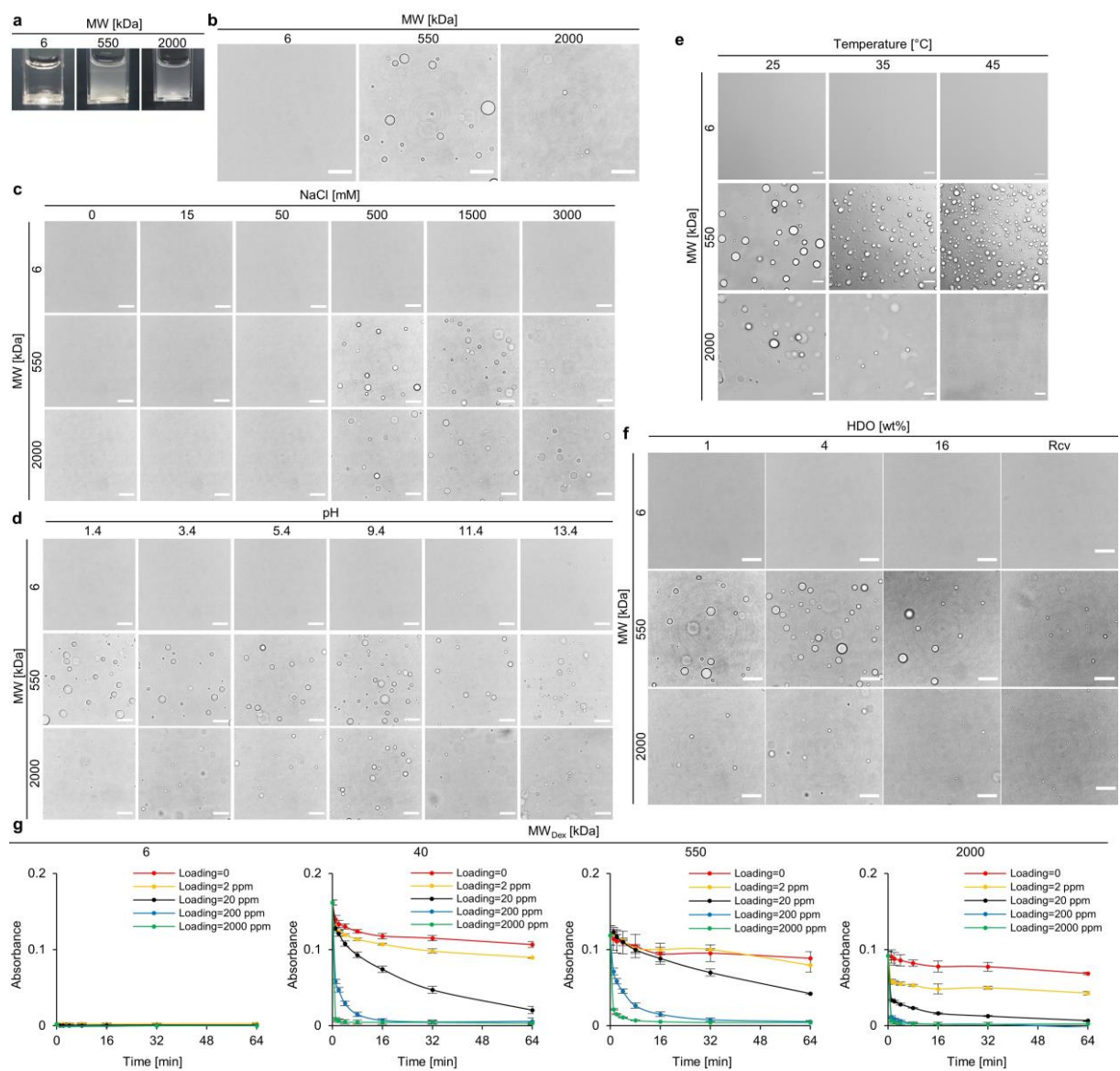

**Supplementary Fig. 6. Imaging LLPS of IPH with different MW. a**, Turbidity change indicates occurrence of LLPS. **b**, Microscale droplets formation. **c**, **d**, **e**, Responsiveness of LLPS to ionic strength (**c**), pH (**d**) and temperature (**e**). **f,** Reversibility of LLPS to HDO. Scale bar=20 µm. **g**, Time-course cleavage of backbone linkage of IPH. Dextranase was loaded at different mass ratios to IPH, namely, $Loading = \frac{m_{E,0}}{m_{S,0}}$. These data indicate multivalency is essential for both initiation and maintenance of LLPS. Scale bar=20 µm.

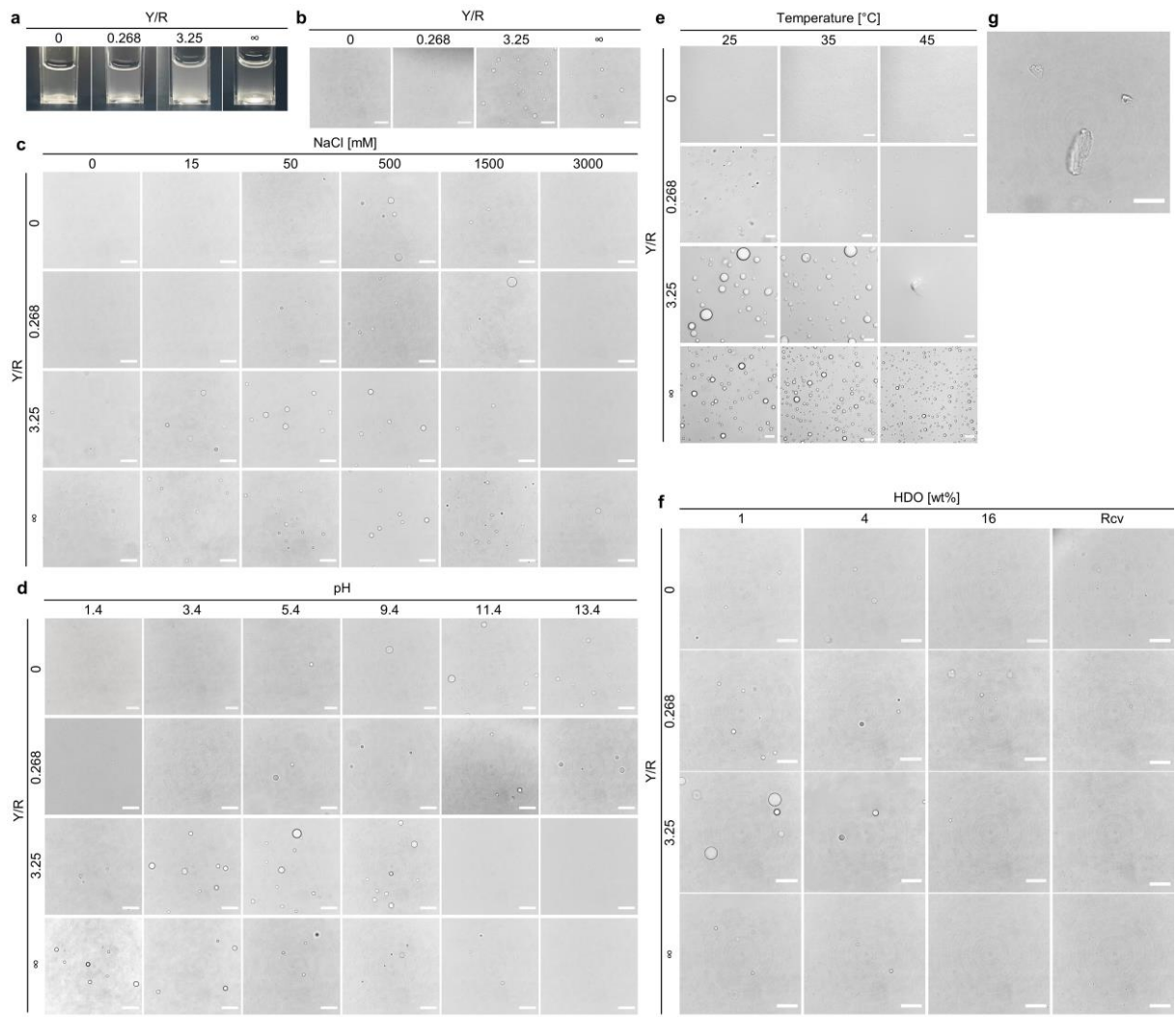

**Supplementary Fig. 7. Imaging LLPS of IPH with different peptide composition. a-f,** Different Y/R ratio. Turbidity change indicates occurrence of LLPS (**a**). Microscale droplets formation (**b**). Responsiveness of LLPS to ionic strength (**c**), pH (**d**) and temperature (**e**). Reversibility of LLPS to HDO (**f**). **g**, Different peptide sequence. Dex40k-CAβACP (DM=37.9 %) was tested for comparison. Scale bars, 20 µm.

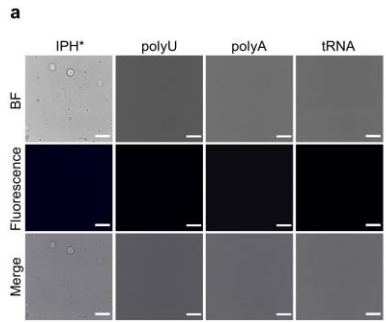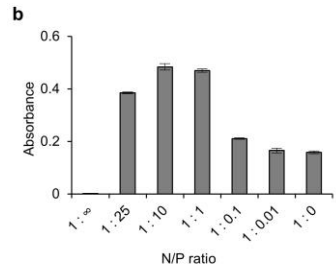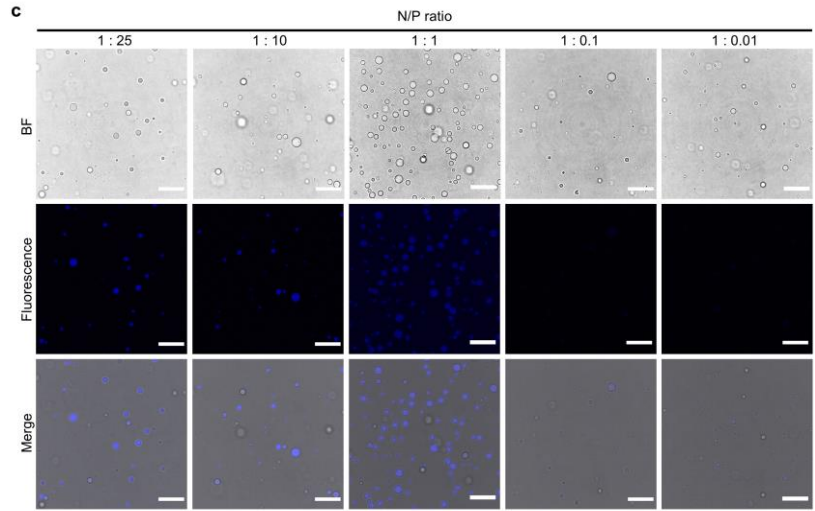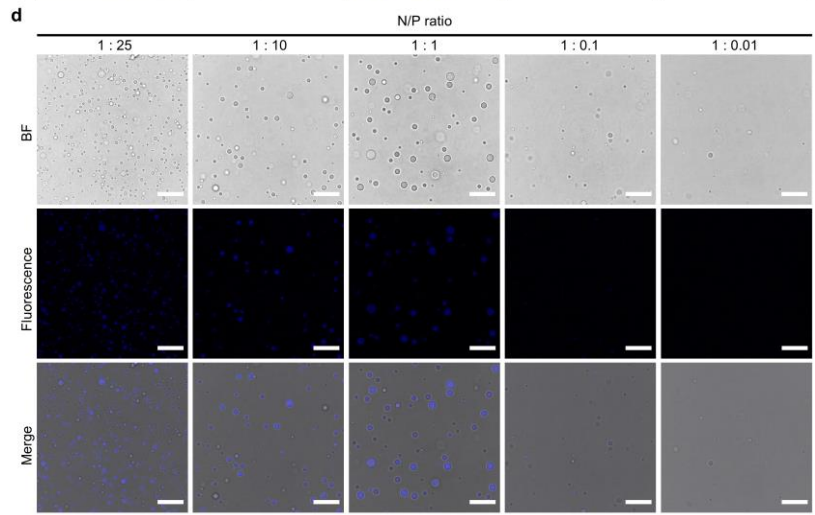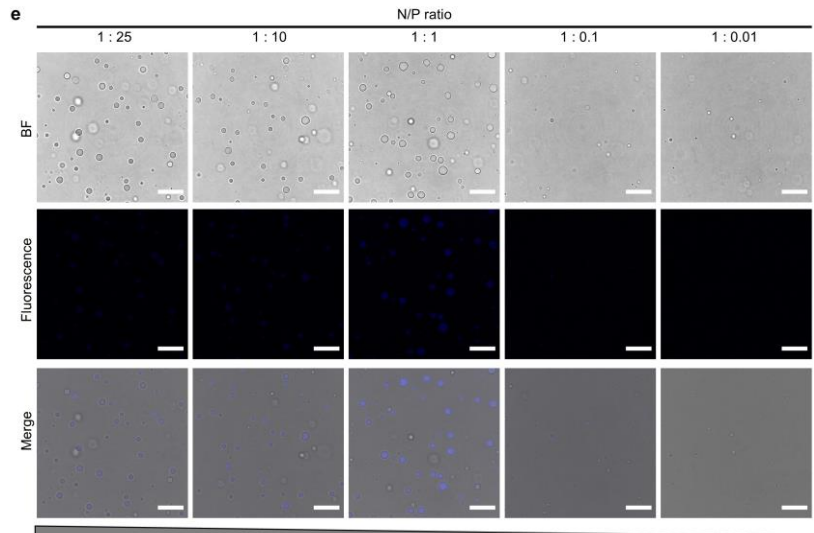

**Supplementary Fig. 8. Complexation of RNA with AMOs. a,** Microscopic image of IPH* (6 μM, N/P ratio=1:0) and RNA solution (0.697 mg/mL, N/P ratio=1:∞). MB shows no evident partitioning in IPH* solution. No evident assembly forms for all the RNAs tested under microscope. **b**, polyU incorporation modulates LLPS propensity of IPH* as indicated by turbidity test. **c-e**, Microscopic image of IPH* complexation and preferential recruitment of polyU (**c**), polyA (**d**), tRNA (**e**). 10 μM methylene blue was incorporated for each set of experiments as the staining agent. Scale bars, 20 μm.

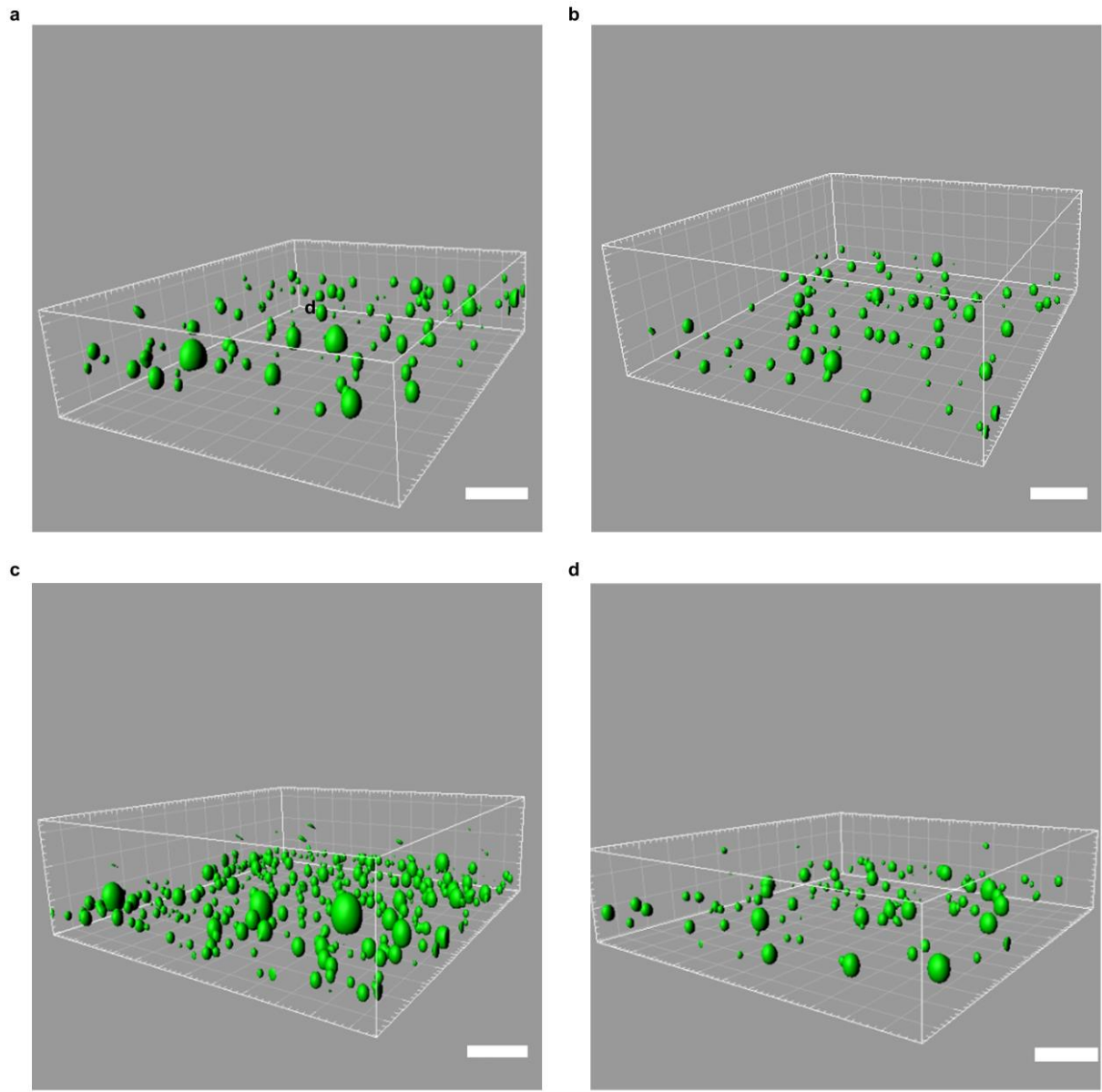

**Supplementary Fig. 9. 3-D rendering. a-d,** 3-D rendering of IPH* (**a**) and IPH* loaded with GFP (**b**), polyU (**c**) and HRP (**d**). Scale bars, 20 µm.

**Supplementary Table 1. Quantification of enrichment of cargoes.**

| Cargo | Loading (μg/mL) | Methods to determine $DC_{\text{dis}}$ | $DC_{\text{dis}}$ (%) | $\phi_{\text{con}}$ (%) | $DC_{\text{con}}$ (%) | $Enrichment$ |
|---|---|---|---|---|---|---|
| GFP | 4 | Autofluorescence intensity assay | 67.2 ± 6.13 | 0.44 ± 0.12 | 7520 ± 1388 | 102 ± 29.8 |
| polyU | 120 | RiboGreen assay | 12.8 ± 1.82 | 0.88 ± 0.325 | 9921 ± 205 | 716 ± 132 |
| HRP | 0.1 | Amplex Red assay | 55.9 ± 6.38 | 0.32 ± 0.026 | 13747 ± 2321 | 246 ± 65.5 |